\documentclass[12pt]{iopart}

\usepackage{iopams,graphicx}  

\newcommand{\wad}[1]{\widehat a^\dagger(#1)}
\newcommand{\wa}[1]{\widehat a(#1)}

\newcommand{\di}{{\rm d}}

\def\wT{{\widehat T}}

\def\wP{{\widehat P}}
\def\wJ{{\widehat J}}

\def\wspt{{\widehat{\cal S}}}

\newcommand{\omegav}{\boldsymbol{\omega}}

\newcommand{\p}{{\rm p}}

\newcommand{\be}{\begin{equation}}
\newcommand{\ee}{\end{equation}}                                                                    
\newcommand{\bea}{\begin{eqnarray}}
\newcommand{\eea}{\end{eqnarray}}

\newcommand{\group}[1]{\relax\ifmmode\mathsf{#1}\else\textsf{#1}\fi}	

\begin{document}

\title[]{Spin and polarization: a new direction in relativistic heavy ion physics}

\author{Francesco Becattini}

\address{Dipartimento di fisica e astronomia, Universit\'a di Firenze\\
and INFN Sezione di Firenze \\
Via G. Sansone 1, I-50019, Sesto Fiorentino (Firenze), Italy}
\ead{francesco.becattini@unifi.it}
\vspace{10pt}
\begin{indented}
\item[]
\end{indented}

\begin{abstract}
Since the first evidence of a global polarization of $\Lambda$ hyperons in relativistic 
nuclear collisions in 2017, spin has opened a new window in the field, both
at experimental and theoretical level, and an exciting perspective. The 
current state of the field is reviewed with regard to the theoretical understanding of the data, 
reporting on the most recent achievements and envisioning possible developments.
The intriguing connections of spin physics in relativistic matter with fundamental 
questions in quantum field theory and applications in the non-relativistic domain
are discussed.  
\end{abstract}

\section{Introduction}

The observation of a global spin polarization of $\Lambda$ hyperons in Au-Au
collisions at $\sqrt{s_{\rm NN}}=200$ GeV \cite{starnature} confirmed, at the 
relativistic and subatomic level, the link between spin and rotation predicted more 
than a century ago and experimentally observed in the Barnett and Einstein-De Haas 
effects \cite{barnett,dehaas}. Besides, it has opened a new and promising direction 
in the field of relativistic heavy ion physics. Since the first positive evidence, 
there has been a considerable progress in understanding this phenomenon both at
theoretical and experimental level. Over the past few years, the experiments have been able 
to confirm the first observations \cite{star1} and demonstrated the capability of 
probing the dependence of spin polarization on momentum \cite{star2}. Besides, the 
measurement of the global polarization of the $\Xi$ hyperon \cite{star3}, in good 
agreement with hydrodynamic predictions, confirmed that this phenomenon is not driven 
by specific hadron-dependent couplings or properties, like in pp collisions, but 
by collective properties of the system, what is often referred to as vorticity-induced 
polarization. Spin polarization has been observed at very low energy \cite{star4,hades} 
and at the highest energy of the LHC \cite{alice1,alice2}. Hence, spin polarization 
is proving to be a practical new degree of freedom which is at our disposal to 
investigate the formation and evolution of the Quark Gluon Plasma (QGP). 

The theory of spin physics in a relativistic fluid also had a significant progress
over the past few years. While the first observation was a successful verification 
of the quantitative prediction of relativistic hydrodynamics and local equilibrium 
\cite{becacsernai,becavort,becakarpe1} spin polarization as a function of momentum 
did not meet the expectations from this model. The discrepancy motivated an intense 
theoretical effort in many 
different directions which has led to a remarkable advance in the understanding 
of spin thermodynamics and kinetics in a relativistic fluid. In fact, because of 
its quantum nature, spin has forced us to reexamine the foundations of relativistic
hydrodynamics and kinetic theory within a quantum field theory framework, widening 
its scope and further clarifying its foundations. Furthermore, as I will discuss
in this work, the study of spin in its native regime - which is relativistic -
may have interesting and highly non-trivial implications for other fields where 
spin is at the focal point, like spintronics \cite{spintro1,spintro2}.

This paper is certainly not a complete overview of the state of the field and of 
all the ongoing developments (interested readers may find a more detailed review 
in \cite{becalisa}, which is quickly getting out of date though). I shall instead 
take the special point of view 
of theory and phenomenology and focus on the recent advances of our theoretical 
understanding of the observed phenomena and on the still open issues. What I find 
mostly exciting in this growing subfield of heavy ion physics is the strong interplay 
between theory, phenomenology and experimental measurements which was typical of 
the golden age of particle physics.

\subsection*{Notation}
In this paper we adopt the natural units, with $\hbar=c=K=1$. The Minkowskian 
metric tensor $g$ is ${\rm diag}(1,-1,-1,-1)$; for the Levi-Civita symbol we use 
the convention $\epsilon^{0123}=1$.
We will use the relativistic notation with repeated indices assumed to be saturated. 
Operators in Hilbert space will be denoted by a wide upper hat, e.g. $\widehat H$.

\begin{figure}
\includegraphics[scale=0.5]{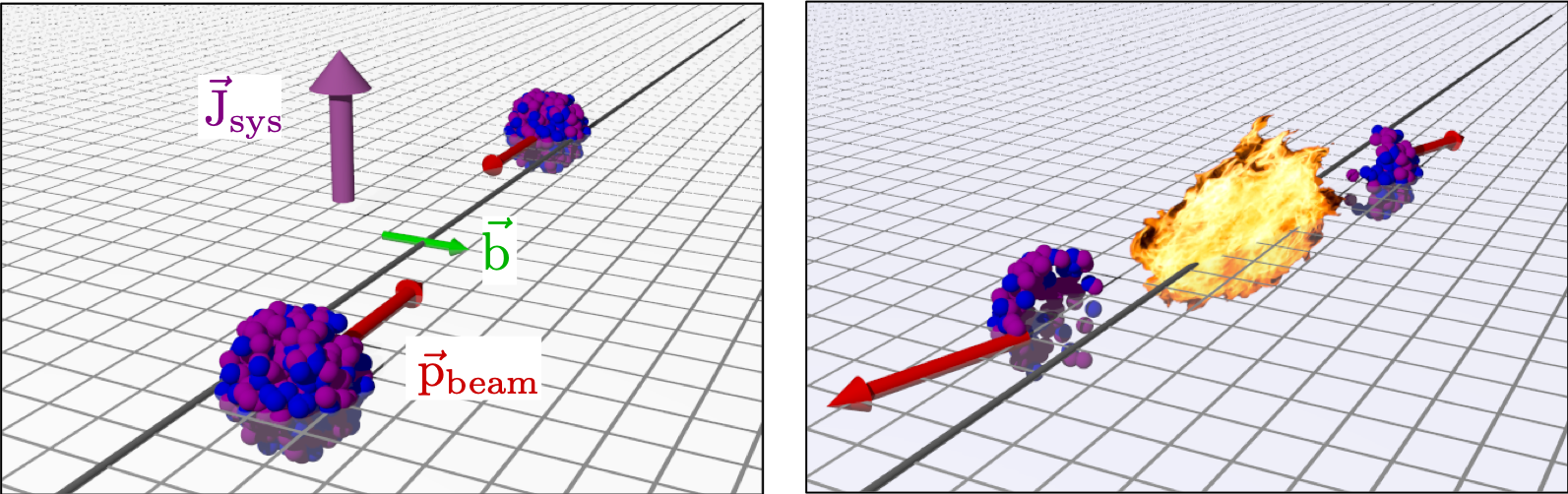}
\caption{Two colliding nuclei at a finite impact parameter ${\bf b}$ produce a 
Quark Gluon Plasma inheriting a fraction of their initial total relative orbital 
angular momentum ${\bf J_{\rm sys}}$ (from ref.~\cite{becalisa}). The angular
momentum vector is perpendicular to the plane containing the momenta of the nuclei
and their impact parameter, the so-called reaction plane.}
\label{sketch}
\end{figure}

\section{Spin and hydrodynamics}

The Barnett effect shows that a rotation of a solid body induces a spin polarization
in the direction of the angular velocity vector \cite{barnett}. It is therefore to
be expected that vorticity in a fluid induces a spin polarization in the direction
of the local vorticity vector. Such an effect has been indeed observed few years ago in 
an experiment using liquid mercury \cite{takahashi}, where a spin current induced 
by a vorticous flow was detected. In heavy ion collisions, two colliding nuclei at 
finite impact parameter (fig.~\ref{sketch}) have, at high energy, a large relative 
angular momentum, whose a relatively large fraction can be inherited by the produced 
QGP in the overlapping region. It is then reasonable to expect that the angular 
momentum gives rise to a finite vorticity in the QGP. The measurements in heavy ion
collisions extend, in a sense, the aforementioned observation to the relativistic 
regime, if the QGP formed in relativistic nuclear collisions is an actual relativistic 
fluid as it is commonly accepted.

Indeed, the hydrodynamic model has become a paradigm for the description 
of the QGP formed in relativistic nuclear collisions. It is by now generally accepted 
that the QGP is a strongly-interacting relativistic fluid which stays close to 
local thermodynamic equilibrium for most of its evolution (the quasi-perfect fluid 
picture), until it breaks up and hadronizes. The success of the statistical-hydrodynamic 
model in reproducing hadron abundances and spectra is seen as an evidence of the 
achievement of local equilibrium prior to the pseudo-phase transition and QGP 
hadronization. Indeed, particle momentum spectra are very well described by the formula: 
\be\label{momspect}
  \varepsilon \frac{\di n_h}{\di^3 \p} = (2S_h+1) \int_\Sigma \di \Sigma_\mu p^\mu 
  \frac{1}{\exp\left[\left( u^\mu(x) p_\mu - \sum_i q_{ih} \mu_i(x)\right) / T(x) \right] 
  \pm 1}
\ee
where $\varepsilon$ is the energy and $p^\mu$ the four-momentum of the particle $h$, 
$S_h$ its spin; $\Sigma$ is the hadronization hypersurface, the sign $+$ applies 
to fermions and $-$ to bosons, and the hydrodynamic input is the temperature $T$,
the relativistic four-velocity $u^\mu$ and the chemical potentials $\mu_i$ coupled 
to the hadronic charges $q_{ih}$. This equation 
is a consequence of local equilibrium and is an excellent leading order approximation 
in the classical limit; subleading terms are dissipative (out-of-equilibrium) 
corrections as well as quantum relativistic equilibrium corrections \cite{becaexact}.

If the local equilibrium picture works well for the momentum, it is a quite natural
idea to extend it to the other, space-time related, degree of freedom of a relativistic 
particle, the spin. Indeed, the counterpart of the eq.~(\ref{momspect}) for the 
mean spin vector, or spin polarization vector, was derived for a relativistic fluid 
at local equilibrium in ref.~\cite{becaspin} and its leading order approximation
in the thermal vorticity (see below) reads:
\be\label{spinvarpi}
 S_\varpi^\mu(p)= - \frac{1}{8m} \epsilon^{\mu\nu\rho\sigma} p_\sigma 
 \frac{\int_{\Sigma} \di \Sigma_\lambda p^\lambda \; n_F (1 -n_F) \varpi_{\nu\rho}}
  {\int_{\Sigma} \di \Sigma_\lambda p^\lambda \; n_F},
\ee
where $n_F$ is a shorthand for the Fermi-Dirac distribution function, i.e. the integrand 
function of (\ref{momspect}) with the sign +:
$$
  n_F = \frac{1}{\exp\left[\left( u^\mu(x) p_\mu - \sum_i \mu_i(x)\right) / T(x) q_{ih}\right] 
  +1}
$$
The thermal vorticity $\varpi$ is defined as:
\be\label{thvort}
  \varpi_{\nu\rho} = -\frac{1}{2} \left( \partial_\nu \beta_\rho - \partial_\rho \beta_\nu \right).
\ee
where $\beta^\mu = (1/T) u^\mu$ is the four-temperature vector, which plays a 
crucial role in relativistic hydrodynamics.
The equation (\ref{spinvarpi}) implies that polarization states are not evenly 
filled and, therefore, the factor $2S_h+1$ in the (\ref{momspect}) is just an 
approximation. Indeed, it is quite an accurate one being the polarization of
the order of few percent. The equation (\ref{spinvarpi}) also demonstrates the peculiar 
feature of spin polarization; unlike the momentum spectrum (\ref{momspect}) 
it depends on the {\em gradients} of the thermo-hydrodynamic fields. Such characteristics 
makes spin polarization a very powerful probe of the hydrodynamic picture, because 
it requires the model to reproduce not just the final configuration of temperature 
and velocity, but its space-time dependence as well, at least near the hadronization 
hypersurface.

It is worth discussing the various terms of the eq.~(\ref{spinvarpi}) by decomposing 
the gradients of the four-temperature vector. Since:
\be\label{fourtemp}
 \partial_\nu \beta_\rho = u_\rho \partial_\nu \left(\frac{1}{T}\right) + \frac{1}{T} 
 \partial_\nu u_\rho 
\ee
and introducing the four-acceleration $A^\mu$ and the four-vorticity vector $\omega^\mu$ 
with the appropriate definitions:
\be\label{accvort}
A^\mu = u \cdot \partial u^\mu  \qquad \qquad
\omega^\mu = \frac{1}{2} \epsilon^{\mu\nu\rho\sigma} \partial_\nu u_\rho u_\sigma
\ee
the antisymmetric part of the tensor $\partial_\nu u_\rho$ can be expressed as
a function of $A$ and $\omega$:
$$
  \frac{1}{2}\left( \partial_\nu u_\rho - \partial_\rho u_\nu \right) = \frac{1}{2} 
   \left( A_\rho u_\nu - A_\nu u_\rho \right) - \epsilon_{\nu\rho\sigma\tau} \omega^\sigma 
   u^\tau
$$
The above decomposition, by using the eqs.~(\ref{thvort}),(\ref{fourtemp}) and 
(\ref{accvort}), makes the integrand of the eq.~(\ref{spinvarpi}):
\be\label{spindeco}
 \epsilon^{\mu\nu\rho\sigma} p_\sigma \partial_\nu \beta_\rho = \epsilon^{\mu\nu\rho\sigma} p_\sigma
   \partial_\nu \left( \frac{1}{T} \right) u_\rho + 2 \, \frac{\omega^\mu u \cdot p 
   - u^\mu \omega \cdot p}{T}
  - \frac{1}{T} \epsilon^{\mu\nu\rho\sigma} p_\sigma A_\nu u_\rho 
\ee
where the $\cdot$ stands for the relativistic scalar product of vectors.
Hence, polarization stems from three contributions: a term proportional to the 
gradient of temperature, a term proportional to the vorticity $\omega$, and a term
proportional to the acceleration. Further insight into the nature of these terms
can be gained by choosing the particle rest frame, where $p=(m,{\bf 0})$ and restoring
the natural units. The eq.~(\ref{spindeco}) then certifies that the spin polarization
vector ${\bf S}^*$ of a spin 1/2 particle, in its rest frame, at some point in 
the fluid, is proportional to the following combination:
\be\label{restframe}
{\bf S}^* \propto \frac{\hbar}{KT} 2 \gamma(\omegav^* - 
(\omegav^* \cdot {\bf u})^* {\bf u}^*/ \gamma^2 c^2) + \frac{\hbar}{KT} {\bf A}^* 
\times {\bf u}^*/c^2 + \frac{\hbar}{KT^2} {\bf u}^* \times \nabla T 
\ee
where, for the occasion, we restored the natural constant $\hbar$ and the Boltzmann
constant $K$. In the eq.~(\ref{restframe}) ${\bf u}^*= \gamma {\bf v}^*,{\bf A}^*,
{\omegav}^*$ are the space components of the four-vectors in
eq.~(\ref{accvort}) and $\gamma = 1/\sqrt{1-v^{*2}/c^2}$; all of the three-vectors
in eq.~(\ref{restframe}) are measured in the particle rest frame. The 
decomposition~(\ref{restframe}) makes it clear what are the thermodynamic 
"forces" responsible for polarization: the first term is the vorticity term, which 
corresponds to the well known Barnett effect \cite{barnett}; the second term is 
an acceleration-driven polarization, and corresponds to the Thomas precession, 
which is expected as the particle is dragged by an accelerated flow; finally, the
last term is a polarization induced by a combination of temperature gradient and 
hydrodynamic flow and should be, to the best of our knowledge, a newly found effect (see
later).

\begin{figure}
\includegraphics[scale=0.5]{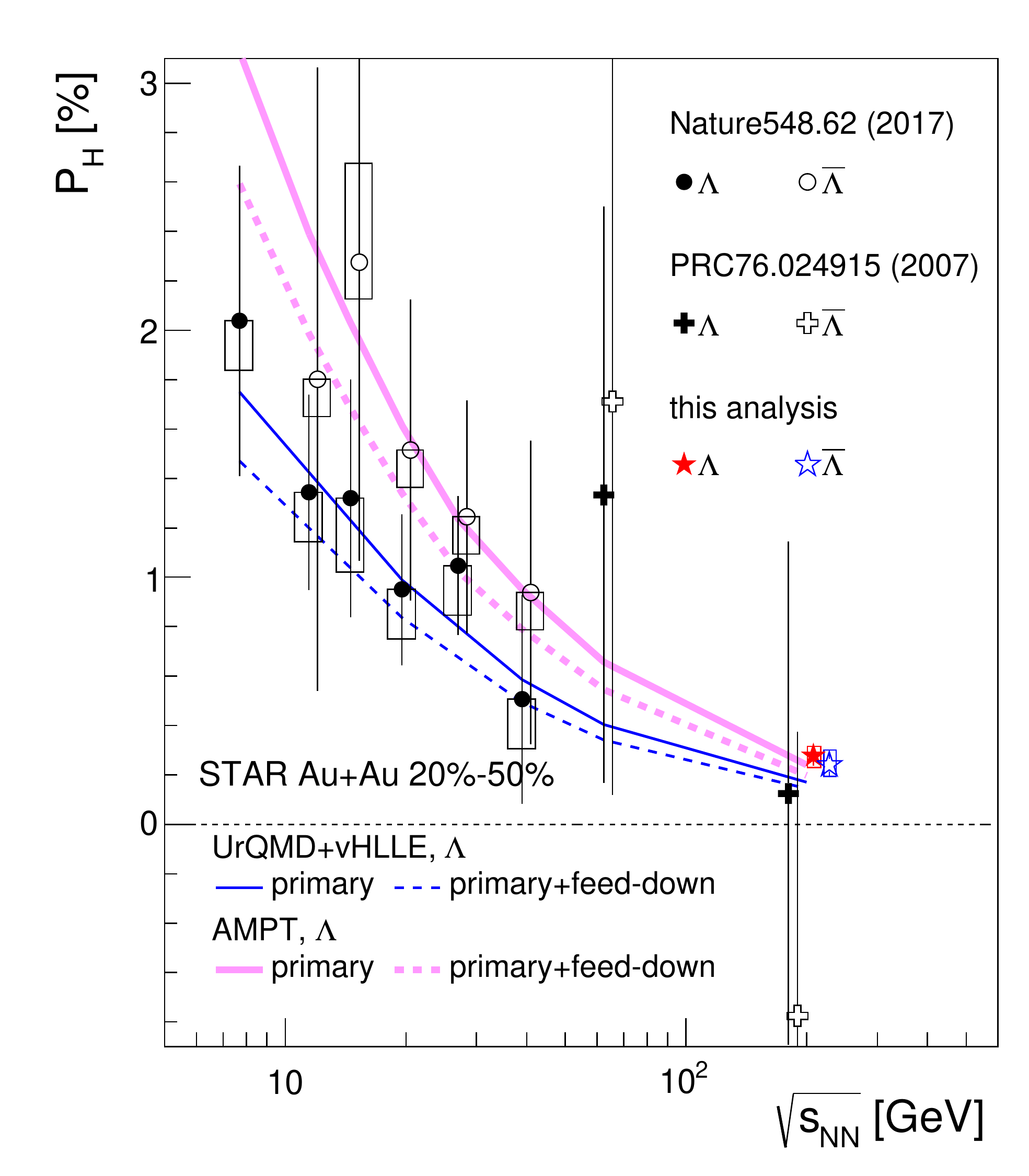}
\caption{Comparison between the measured global polarization of $\Lambda$ hyperons
in relativistic heavy ion collisions at a nucleon centre-of-mass energy of 200 GeV
and the predictions of equation~(\ref{spinvarpi}. Red and blue lines refer to 
different models to calculate thermal vorticity field at the hadronization (from 
ref.~\cite{star1}).}
\label{starfig}
\end{figure}

Indeed, the eq.~(\ref{spinvarpi}) successfully reproduced the global spin polarization, 
integrated over momentum (see fig.~\ref{starfig}), which yields a vector directed
along the initial angular momentum of the collision (see fig.~\ref{sketch}).
Yet, the same equation failed to reproduce the spin dependence on momentum. 
The components of the spin polarization vector of the $\Lambda$ hyperons along
the total angular momentum direction and along the beam line (longitudinal polarization), 
as a function of the hyperon azimuthal angle were measured by the STAR collaboration
~\cite{star2,niida} for hyperons around zero rapidity and found to strongly disagree 
with the predictions of the formula (\ref{spinvarpi}) (see e.g. the discussion in 
ref.~\cite{becalisa}). 

This puzzle has lasted several years. Recently, however, a new local equilibrium 
term which is linear in the gradients, hence comparable with the thermal vorticity 
contribution, was found out with two independent derivations \cite{becashear,yiyin}
(and later confirmed in the study of ref.~\cite{dly}). The formula in 
ref.~\cite{becashear} reads:
\begin{equation}\label{spinxi}
S_\xi^\mu(p)= -\frac{1}{4m} \epsilon^{\mu\rho\sigma\tau} \frac{p_\tau p^\lambda}{\varepsilon}
 \frac{\int_{\Sigma} \di \Sigma_\nu p^\nu \; n_F (1 -n_F) 
 \hat{t}_\rho\xi_{\sigma\lambda}}{\int_{\Sigma} \di \Sigma_\nu p^\nu \; n_F}
\end{equation}
where $\hat{t}$ is the time direction in the QGP frame and $\xi$ is the symmetric 
derivative of the four-temperature, defined as {\em thermal shear} tensor:
\be
  \xi_{\mu\nu} = \frac{1}{2} \left( \partial_\mu \beta_\nu + \partial_\nu \beta_\mu \right) .
\ee
Indeed, the formula (\ref{spinxi}) involves some geometrical approximations of the 
freeze-out hypersurface; in the derivation in ref.~\cite{yiyin} a different 
approximation scheme was used and, as a result, the $\hat t$ direction is replaced 
by the local four-velocity \footnote{It should also be noted that the eq.~(\ref{spinxi}) 
is weakly dependent on the pseudo-gauge choice, even if the spin potential differs 
from thermal vorticity, see Section~\ref{spintens}. Particularly, it was shown in 
ref.~\cite{becashear} that for spin 1/2 particles, the right hand side is the 
same for both the canonical and the Belinfante pseudo-gauges.}.
Hence, altogether, the spin polarization vector at local equilibrium is the sum 
of the right hand sides of the eqs.~(\ref{spinvarpi}) and (\ref{spinxi}):
\be\label{spinpol}
  S^\mu_{\rm LE} = S_\varpi^\mu(p) + S_\xi^\mu(p)
\ee
The inclusion of (\ref{spinxi}) proved to be able to consistently reduce the 
discrepancies \cite{baochifu} and indeed fully restore the agreement with the 
data at top RHIC energy at $\sqrt s_{\rm NN} = 200$ GeV, for both the components 
of polarization, once the temperature gradients are removed to take into account 
that the hadronization hypersurface is $T={\rm const}$ \cite{becaisothermal}.
This latter assumption is a very reasonable one at very high energy, where the chemical potentials
basically vanish and the only thermodynamic parameter governing the hadronization
transition is temperature. In conclusion, even though this picture needs to be 
confirmed by more studies (see e.g. \cite{florkshear} and \cite{ryu2}) and tests 
at different energies,  
it is fair to conclude that the term (\ref{spinxi}) is a likely solution of the 
discrepancies between the azimuthal dependence of polarization and the local 
equilibrium hydrodynamic predictions at very high energy. In general, the term 
(\ref{spinxi}) is to be introduced in all calculations of spin polarization. 

It should be stressed that the additional term (\ref{spinxi}) is not a dissipative
correction like the viscous stress tensor for the stress-energy tensor. In fact, 
this term is a quantum correction to the local equilibrium, hence it is non-dissipative,
which vanishes at global equilibrium where $\xi=0$. The non-dissipative nature of this
term is also revealed by the absence of any dynamical transport coefficient in the
right hand side of eq.~(\ref{spinxi}); it can be seen that it just depends on free-particle
quantities, just like the (\ref{spinvarpi}). As such, this term is an unexpected 
contribution, even more so as the only non-relativistic known formula relates 
the spin vector to the angular velocity \cite{becalisavolo} (in this respect, the 
equation (\ref{spinvarpi}) appears to be a more typical relativistic extension of 
a non-relativistic formula). In fact, the eq.~(\ref{spinxi}) has a non-trivial 
non-relativistic limit which is a single term involving the gradient of the temperature 
field \cite{becashear}:
\be\label{nonrel}
  {\bf S}_\xi = \frac{1}{8} {\bf v} \times 
   \frac{\int \di^3 {\rm x} \; n_F (1-n_F) \nabla  \left( \frac{1}{T} \right)}
  {\int \di^3 {\rm x} \; n_F } 
\ee
where ${\bf v}$ is the velocity of the particle. Interestingly, the eq.~(\ref{spinvarpi}) 
features the same term in its non-relativistic limit by going to the local rest
frame of the fluid where $u=(1,{\bf 0})$, so that the non-relativistic 
limit of (\ref{spinpol}) includes twice the right hand side of the eq.~(\ref{nonrel}). 
This equation predicts that a particle travelling in a medium with a temperature 
gradient will get polarized, thus giving rise to a tiny magnetic field and a spin 
current along its motion \footnote{In this case, we define spin current as the 
motion of polarized particles. The proper definition of spin current in quantum 
field theory requires some care.}. Thermal spin effects have been predicted and 
observed in condensed matter \cite{adachi} but it is not clear whether this particular 
one has been observed yet. 

There are many more measurements that can be done to further test the hydrodynamic-local 
equilibrium picture. For instance, in ref.~\cite{xia}, a proposal of measuring the 
polarization dependence on the azimuthal angle at finite rapidity instead of midrapidity
to probe the vorticity field. To date, the spin polarization has been measured
by the experiments also as a function of transverse momentum, centrality and rapidity
\cite{star1,star2,alice2}.  
An interesting study of the hydrodynamic predictions on these dependences has been 
recently carried out \cite{ryu}, where the new term (\ref{spinxi}) and a similar 
local equilibrium term proportional to the gradient of $\mu_B/T$ \cite{yiyin2} (which 
is, in general, more relevant at low energy, see below) were included. Therein, 
the authors showed that the effect of the term in 
eq.~(\ref{spinxi}) on the global polarization (integrated over all momenta) is almost 
negligible, thus confirming that the agreement between global polarization and the 
predictions based solely on the term (\ref{spinvarpi}) is not accidental. Furthermore, 
the authors systematically studied the dependence of the hydrodynamic predictions 
on the initial longitudinal flow velocity, showing that the spin polarization can 
be used as a major probe of the initial hydrodynamic conditions, which is itself 
a remarkable fact. 

Finally, there have been very interesting studies of polarization at very low energy. 
The experiment STAR has been able to push the lowest energy polarization measurement
down to $\sqrt{s_{\rm NN}}=3$ GeV in order to test the typical prediction of the 
hydrodynamic model of an increasing polarization as energy decreases \cite{becakarpe1}. 
It is not yet clear if the same hydrodynamic model which is used at high energy 
can be applied at such a low energy, however there are calculations 
\cite{ivanov,xghlow1,xghlow2,liaolow} based on hydrodynamics and other models 
predicting the appearance of a peak of global polarization as a function of 
collision energy. The magnitude measured by 
the collaboration STAR \cite{star4} seems to favour the three-fluid model \cite{ivanov}, 
however it should be stressed that none of the above calculations included the 
spin-shear term (\ref{spinxi}). In fact, it has not been determined if the spin-shear 
term can sizeably contribute to the polarization at low energy. However,
a recent study showed that at low energy, where the baryon chemical potential is large, the 
effect of the local equilibrium term proportional to $\partial (\mu_B/T)$ (dubbed 
as Spin Hall Effect \cite{adachi} in ref.~\cite{yiyin2}) may play an essential role, 
with an almost a factor 2 increase of the global polarization magnitude \cite{yiyin3}.
Another reason for the interest of low energy polarization studies is the apparent
splitting between $\Lambda$ and $\bar\Lambda$ polarization (see fig.~\ref{starfig}), 
for which the electro-magnetic field is possibly responsible \cite{becalisa,liaomag}. 

\subsection{Beyond local equilibrium}

What if deviations from local equilibrium picture eventually hold? If the discrepancies 
between predictions and data are significant but small (what seems to be a fair 
conclusion after the inclusion of the new term (\ref{spinxi}), as discussed), 
then it can be argued that dissipative corrections could cure them. 
The theoretical problem would be alike to the calculation of the dissipative corrections 
of the particle momentum distribution (\ref{momspect}) with viscous terms, which is 
a well known problem in the hydrodynamic modelling of heavy ion collisions (see 
e.g. \cite{molnar}). The main difficulty, in the spin 
case, is the identification of the involved dissipative coefficients, whether they
are the known ones (shear and bulk viscosity, thermal conductivity) associated to
the stress-energy tensor, or new ones which have not yet been classified. Indeed, 
an expression has been derived for the dissipative corrections of the spin polarization 
vector of a system of massless fermions \cite{shigale} (see also ref.~\cite{hidyang})
with terms proportional 
to the viscous part of the stress-energy tensor and to the diffusion currents.
Recently, an expression of the dissipative corrections for massive fermions, involving
several new coefficients, has been obtained based on relativistic kinetic theory 
with spin \cite{noradirk}. Apparently, these terms have been derived with a special 
choice of the spin tensor (see below), and it is not yet clear if they are independent
thereof, what promises to be a really intriguing question.

If, on the other hand, large discrepancies will appear, local equilibrium 
could not be longer considered a good approximation. In this case, to get a better
description, there are two possible options: extending the notion of local equilibrium 
by introducing a spin potential (which will be discussed in detail in Section~\ref{spintens}) 
or a full non-equilibrium approach, such as a quantum kinetic description. Both 
these cases imply a separation of scales between the spin relaxation time and the 
typical interaction time. 

The kinetic theory with spin is an extension of relativistic kinetic theory and 
had a remarkable 
progress over the past few years, with many contributors. The most suitable tool 
is quantum kinetic theory with collisions \cite{qktfrank1,qktfrank2,yanghh,zwang,yang21,
hidaka}, which is very useful to obtain the constitutive equations of spin hydrodynamics \cite{sw} 
and other theoretical insights. Yet, it is not clear if the kinetic approach, based 
on the picture of quasi-free colliding particles \cite{zhang}, requiring the mean 
free path to be much larger than the thermal wavelength, is phenomenologically 
viable for the QCD plasma near the pseudo-critical temperature.
The very low value of the viscosity/entropy density ratio entails comparable magnitudes
for the mean free path and thermal wavelength (the so-called strongly-interacting
QGP), implying that the collisional picture is questionable. Nevertheless, if the 
spin relaxation time is much longer than the typical interaction time scale, perhaps 
a kinetic description limited to spin-changing interactions could be still a good 
approximation.

\section{The spin tensor, the spin potential and spin hydrodynamics}
\label{spintens}

Right about the same time evidence was found of polarization in relativistic heavy 
ion collisions, the question arose about if and how including this new degree of
freedom in the hydrodynamic model of the QGP. From this problem a new line of research 
has sprung, the so-called relativistic spin hydrodynamics, with many contributions.

The basic idea of relativistic spin hydrodynamics (see e.g. \cite{florfri}) is to 
describe the dynamics of the relativistic fluid with an additional tensor of rank 3, 
the spin tensor ${\cal S}^{\lambda,\mu\nu}$, which is anti-symmetric in the last 
two indices, besides the familiar stress-energy tensor $T^{\mu\nu}$. The spin tensor 
fulfills a continuity equation dictated by the conservation of angular-momentum:
\be\label{spineom}
  \partial_\lambda {\cal S}^{\lambda,\mu\nu} = T^{\nu\mu} - T^{\mu\nu}
\ee
(the stress-energy tensor not being necessarily symmetric) which is the additional
fundamental equation of relativistic spin hydrodynamics. 

There is a fundamental issue though, namely whether the spin tensor has a real 
physical meaning. This is a sort of ever-returning highly-debated question in several 
fields, like e.g. proton spin studies \cite{leaderlorce}, and it has to do with the 
possibility of a unique separation of the orbital and the spin angular momentum 
in relativity or, tantamount, the definition of a unique stress-energy tensor. 
Ultimately, this question is related to a crucial feature of the conserved currents 
in quantum field theory, the so-called {\em pseudo-gauge invariance}, that is the 
possibility of changing the spin and the stress-energy tensors at the same time 
with a suitable linear transformation preserving the continuity equations and the 
total energy-momentum and angular momentum \cite{hehl,becatinti}, which are obtained
by integrating the densities over an arbitrary 3D space-like hypersurface:
\bea
 \wP^\nu &&= \int \di \Sigma_\mu \; \wT^{\mu\nu}(x) \\ \nonumber
 \wJ^{\lambda\nu} && = \int \di \Sigma_\mu \; \left( x^\lambda \wT^{\mu\nu}(x)
- x^\nu \wT^{\mu\lambda}(x) + \wspt^{\mu,\lambda\nu} \right) \nonumber
\eea
Otherwise stated, the stress-energy tensor and the spin tensor, in relativistic
quantum field theory, are much alike to gauge potentials, which are physically 
unmeasurable. This is confirmed by the fact that the operator $\widehat S^\mu(p)$,
whose mean value is the spin polarization vector measured in the experiment, can
be expressed in terms of creation and annihilation operators of the quantum field
and does not depend on the spin tensor \cite{becalibro}:
$$
 \widehat S^\mu(p) = \sum_i [p]^\mu_i D^S(J_i)_{rs} \wad{p}_r \wa{p}_s
$$
In the above equation, $D^S(J_i)$ are the representation matrices of the generators 
of rotation for the spin $S$, $[p]$ is the so-called standard Lorentz transformation 
taking the four-vector $(m,{\bf 0})$ to the four-momentum $p$ and the $\wa{p}_s$ are the 
annihilation operators of a particle with four-momentum $p$ and spin state $s$.

The spin tensor, which is charge-conjugation even, contains information about the 
spin density and the spin current within matter at some point, but it should be 
emphasized that it is necessary only in a quantum relativistic theory with 
anti-particles, to describe a situation where both particles and antiparticles 
can be polarized in the same direction. Indeed, in the ordinary matter, without 
anti-particles, a finite spin density can be described by the magnetization tensor 
(e.g. in the Dirac theory $\bar\Psi [\gamma^\mu,\gamma^\nu] \Psi$) which is 
charge-conjugation odd. This means that the Barnett effect, i.e. magnetization by
rotation in ordinary matter, is completely insensitive to the spin tensor and 
does not really break pseudo-gauge invariance.

It appears to be somewhat surprising that the stress-energy tensor is a sort of 
gauge field, but it should be kept in mind that, strictly speaking, the experiments 
cannot actually measure densities in space. In fact, they can only measure 
momentum spectra or spin polarization as a function of momentum. Therefore, all 
the information on the local state of the fluid, in terms of the hydrodynamic-thermodynamic
fields, is inferred through relations like the eq.~(\ref{momspect}) or the 
eq.~(\ref{spinvarpi}). These fields, particularly the four-temperature field $\beta(x)$,
are defined by enforcing local thermodynamic equilibrium conditions, which turn out to
be pseudo-gauge dependent \cite{bfs}, as it will become clear below. Therefore, 
the densities, the currents and the thermodynamic fields are pseudo-gauge dependent, 
meaning that they are physically objective only up to small, yet finite quantum terms.

The most apparent manifestation of pseudo-gauge invariance is the possibility to
make the spin tensor vanishing altogether, the so-called Belinfante pseudo-gauge. 
This choice seems to imply that the very notion of "spin density" (beware the difference
with magnetization density) is unphysical but in fact it means that in a relativistic 
fluid particles and anti-particles cannot be polarized with the same sign without 
a thermal vorticity, i.e. without a rotation \cite{bfs,stephanov}.
In other words, in the Belinfante pseudo-gauge, a relativistic fluid cannot present 
itself in a state where particles and anti-particles are polarized in the same direction
without rotation, or better, with vanishing thermal vorticity. To describe such a
situation, the prepared quantum state needs to break pseudo-gauge invariance, 
and indeed a density operator or a superposition of states can - at least in principle - 
be constructed which explicitely depends on the pseudo-gauge couple
$(T^{\mu\nu},{\cal S}^{\lambda,\mu,\nu})$ \cite{bfs,sw,becaqm2019}.

In a kinetic-based approach, where particles are the fundamental objects, a 
situation with particles and anti-particles polarized in the same direction 
without an associated thermal vorticity pertains to a system where the spin relaxation 
time is longer than momentum relaxation time scale \cite{bfs,stephanov}, so that 
spin can be considered as an independent hydrodynamic slow mode \cite{stephayin}
which is not locked to thermal vorticity. In this case, local equilibrium of angular 
momentum density requires a new intensive thermodynamic quantity, an anti-symmetric 
tensor called the spin potential $\Omega_{\mu\nu}$ 
\footnote{This quantity is called spin {\em chemical} potential in some papers. 
I find the adjective "chemical" inappropriate because it has nothing to do with 
the internal quantum numbers and it is in fact common to all particle species.},
which is unnecessary in the Belinfante pseudo-gauge. The spin potential couples 
to the spin tensor in the exponent of the local equilibrium density operator 
\cite{bfs}, but the problem is which spin tensor (and, consequently, which
stress-energy tensor) among all the possible pseudo-gauge choices, is to be chosen
for the implementation of the local angular momentum density constraint. One can
appeal to an external extended gravitational theory \cite{stephanov} or to other 
arguments \cite{florkowski}, but in essence the choice is free in quantum field 
theory in flat spacetime.

The final goal of spin relativistic hydrodynamics is to determine the spin potential
$\Omega_{\mu\nu}$ at the particle freeze-out with given initial conditions. Hence, 
spin relativistic hydrodynamics is an extension of relativistic hydrodynamics with 
six additional fields to be evolved besides the four-temperature $\beta$ and chemical 
potentials. The effective number of independent dynamical variables of the spin potential 
might be reduced \cite{stephanov} if additional symmetries of the spin tensor are present, like
for the canonical Dirac spin tensor which is completely anti-symmetric.
The conservation equation of the mean value of the stress-energy tensor:
$$
   \partial_\mu T^{\mu\nu}[\beta,\zeta,\Omega] = 0 
$$
(the square brackets stand for a {\em functional} dependence, that is on the function
and all the derivatives at the point $x$) is supplemented by the continuity equation
of the mean value of the spin tensor (\ref{spineom}) which makes it possible to 
find a solution for all the fields once the initial and boundary conditions are 
provided.

The spin potential enters the spin polarization vector expression, modifying 
the eq.~(\ref{spinpol}) \cite{buzzegoli,liuxgh} with an additional contribution, 
for the canonical spin tensor:
\be\label{spinomega}
\Delta S^\mu(p) = \frac{\epsilon^{\lambda\rho\sigma\tau}\hat t_\lambda 
 \left(p_\tau p^\mu - \delta^\mu_\tau m^2 \right)}{m \varepsilon}
 \frac{\int_{\Sigma} \di \Sigma_\nu p^\nu \; n_F (1-n_F) 
  (\varpi_{\rho\sigma}-\Omega_{\rho\sigma})}
  {\int_{\Sigma} \di \Sigma_\nu p^\nu \; n_F},
\ee
which makes it apparent that a non-vanishing spin polarization is possible even 
with $\varpi=\xi=0$ (see discussion above).

The topic of spin tensor and spin hydrodynamics has attracted a lot of interest among 
theorists, because of the enjoyable work of deriving the constitutive equations and 
the dissipative terms of the spin tensor \cite{hattori,liao,hu,florkdiss,yee,
stephanov}. It is of course a very interesting subject, and a possible solution of 
a disagreement between data and the predictions of the eq.~(\ref{spinpol}) but the 
price to be payed, from the point of view of phenomenology, is high. 
Besides the general problem of picking a particular spin tensor to be coupled to 
the spin potential, the introduction of six additional fields $\Omega_{\mu\nu}(x)$ 
requires the specification of their initial conditions, which are essentially 
unknown. The only observable which is sensitive enough to the spin potential is 
the spin polarization, through the above equation (\ref{spinomega}).
Therefore, any measurement of spin polarization would serve to adjust the initial
conditions of the spin potential rather than to test a theoretical prediction; put it simply,
it would be a great loss of predictive power of the hydrodynamical model. This
is indeed a real possibility, but a very disappointing one. So far, there is no
evidence that a spin potential different from thermal vorticity is necessary to 
describe the data, but further tests and more accurate measurements may lead to
a different conclusion. 

\section{Spin alignment}

The local equilibrium picture predicts that all sort of particles come out polarized from
the hadronizing plasma, with a polarization depending on the hydrodynamic fields
and their spin \cite{becalisavolo}. The mean spin polarization vector $S^\mu$, in 
practice, can be measured in heavy ion experiments only for weakly decaying hyperons, 
such as $\Lambda$, $\Xi$ and $\Omega$. For spin 1 mesons, which do not decay weakly, 
a polarization-related quantity which is accessible through the decay process 
is the so-called alignment:
$$ 
\Theta_{00} -1/3
$$
where $\Theta_{00}$ is the diagonal component of the spin density matrix, $0$ being 
the eigenvalue of the third componentof the spin operator $\widehat S_z(p)$. The 
possibility of detecting alignment in peripheral heavy ion collisions attracted 
the attention \cite{liang} almost as early as polarization \cite{voloshin,xnw}.

The alignment has been measured for the ${\rm K}^*$ and the $\phi$ vector mesons at 
$\sqrt s_{\rm NN} = 200$ GeV \cite{singha} and $\sqrt s_{\rm NN} = 2.76$ TeV \cite{mohanty} with 
different outcomes: a consistently negative value for the ${\rm K}^*$ at low $p_T$ at 
both energies, whereas, for the $\phi$, a slightly positive value at the lower energy 
and a negative value at low $p_T$ at the higher energy. For all of these measurements
the magnitude of the alignment (especially for the ${\rm K}^*$) is apparently too large 
to be accomodated within a local equilibrium picture, which predicts a quadratic dependence 
on thermal vorticity \cite{becapicci} and, most likely, on the thermal shear too. 
Besides, it is difficult to explain the change of sign of the $\phi$ alignment from
200 to 2760 GeV. Several theoretical attempts have been made to reproduce these measurements
which do not rely on local equilibrium, but on quark coalescence \cite{xiaali}, 
collective effects \cite{sheng} or, more recently, on calculations of quantum 
kinetic theory in weakly coupled QCD \cite{yangmuller}. 

In my view, spin alignment is still an unsolved problem. It should be pointed 
out that the change of sign of $\phi$ alignment between the two explored energies 
is itself a puzzling evidence, difficult to reproduce in most models. 

\section{What is the spin good for?}

As I have tried to emphasize, the study of spin-related phenomena in relativistic 
matter is a commendable effort for relativity is a more general and natural viewpoint 
for spin and it can lead to a deeper understanding of the non-relativistic phenomena, 
which are of course most likely to have practical applications. Still, spin may 
have other fruitful applications in the very field of relativistic heavy ion physics.
 
Indeed, as spin is the newest chapter for this field, thus far most efforts 
went and are still going into checking that spin measurements are understood within a
commonly accepted theoretical framework, what has been the subject of previous sections. 
As the theoretical picture consolidates, the question arises whether spin can be 
used as a tool to attack fundamental unanswered or partially answered questions 
on the QCD matter, the QGP and its properties and evolution. 
Indeed, there have been interesting proposals in this respect and I would like to
mention three of them. 

A first proposal is to use the measurement of helicity - that is the projection 
of the mean spin vector in the particle rest frame onto the momentum direction -
to detect local parity violation in the QCD matter, the long-sought phenomenon 
with the Chiral Magnetic Effect (CME) \cite{cme}. The idea of revealing local parity 
violation with the helicity correlation of $\Lambda$ hyperons was put forward
in ref.~\cite{finch} and has been recently proposed again in ref.~\cite{helicity}
where a formula was found connecting the helicity to the axial chemical potential
$\mu_A$ developed as a consequence of the local parity violation induced at high 
temperature \cite{kharzpisa}:
\be\label{spinaxial}
S^\mu_{\chi}(p) \simeq \frac{g_h}{2} \frac{\int_\Sigma \di\Sigma_\lambda p^\lambda \; 
(\mu_A/T) n_F \left(1-n_F \right)}{\int_\Sigma \di\Sigma_\lambda p^\lambda \; 
n_F} \frac{\varepsilon p^\mu- m^2\hat t^\mu}{m \varepsilon}
\ee
where $\hat t$ is the unit vector in the QGP center-of-mass frame, as before, 
and $g_h$ is the axial charge of the $\Lambda$.
The detection of local parity violation through the measurement of the helicity 
azimuthal correlation function of pairs of $\Lambda$ hyperons does not require 
a coupling with the electro-magnetic field and it is thus complementary to the 
CME. The feasibility of such a measurement, in terms of statistics, experimental
backgrounds and errors, is still under investigation.

A second possibility offered by spin measurements is to investigate the energy
loss of highly energetic partons, i.e. jets, in the QGP. It has been pointed out 
by Serenone et al.~\cite{lisajet} that a fast parton losing energy within the plasma 
would induce a vorticity field (more generally, a gradient of the hydrodynamic fields). 
Such a non-trivial field pattern can be probed by a measurement of the polarization 
of the emitted $\Lambda$ hyperons as a function of their angular distance from 
the jet axis. The authors also showed that the amplitude of such induced polarization
is very sensitive to the shear viscosity/entropy density ratio $\eta/s$. 
 
A third, very recent idea, is to use spin polarization to search for the QCD 
critical point. It has been observed by the authors of ref.~\cite{singhalam}
that, in the proximity of the critical point, the change in the equation of state
and the scaling behaviour of the transport coefficients, affects the thermal vorticity
field to a sizeable extent, hence the final polarization of hyperons. Particularly,
it turns out that the dependence of $S^*_J = {\bf S}_0 \cdot {\bf \hat J}$ spin component 
along the total angular momentum on the rapidity of the $\Lambda$ is markedly 
different if the critical point is there, with a relatively strong suppression at 
midrapidity. As the magnitude of polarization at lower energy is larger than at 
high energy, a measurement is likely to be feasible with a sufficiently high 
statistics of $\Lambda$ hyperons at the forthcoming heavy ion facilities NICA and FAIR.

\section{Conclusions}

Spin has opened a new window in the physics of the QGP and relativistic heavy ion
collisions. As I have tried to emphasize, its sensitivity to the gradients of the
thermo-hydrodynamic fields make it a powerful probe of the hydrodynamic
model of the plasma and of its initial conditions. Besides, the potentialities
of the study of spin in this field with respect to a variety of phenomena and other
fields such as spintronics, are still to be fully explored and understood.

While I am writing, theory and experiments are still in a quickly developing stage, with 
a fast and intense mutual and beneficial interaction. Over the past year, the progress
in phenomenology has been remarkable and, in spite of the high level of 
accuracy which is demanded in the spin-related calculations, several groups in the 
world proved to be able to make consistent numerical simulations. At the same time, 
the experiments proved to be able to carry out more and more differential measurements. 
In the future, a further increase of collected statistics will make it possible 
to test even more detailed observables, such as spin-spin correlations \cite{elfner} 
which will probe the hydrodynamic, as well as other models, in more depth.

\section*{Acknowledgments}

I gratefully acknowledge interesting discussions and very useful suggestions 
by J. P. Blaizot, X. G. Huang, J. F. Liao, D. Rischke, S. Voloshin, N. Weickgenannt 
and D.L. Yang.
 


\end{document}